\documentclass[%
 aip,
 amsmath,amssymb,
 reprint,%
]{revtex4-1}

\usepackage{graphicx}
\usepackage{dcolumn}
\usepackage{bm}

\usepackage[utf8]{inputenc}
\usepackage[T1]{fontenc}
\usepackage{mathptmx}
\usepackage{etoolbox}

\usepackage{xcolor}

\makeatletter
\def\@email#1#2{%
 \endgroup
 \patchcmd{\titleblock@produce}
  {\frontmatter@RRAPformat}
  {\frontmatter@RRAPformat{\produce@RRAP{*#1\href{mailto:#2}{#2}}}\frontmatter@RRAPformat}
  {}{}
}%
\makeatother
\begin{document}

\preprint{AIP/123-QED}

\title[Quantum-on-chip]{Quantum photonics on a chip}
\author{A. Katiyi}
\affiliation{ 
School of Electrical and Computer Engineering, Ben-Gurion University of the Negev, Beer-Sheva 8410501, Israel
}%
\author{A. Karabchevsky}%
 \email{alinak@bgu.ac.il}
\affiliation{ 
School of Electrical and Computer Engineering, Ben-Gurion University of the Negev, Beer-Sheva 8410501, Israel
}%

 \homepage{https://www.alinakarabchevsky.com/.}
\affiliation{%
Department of Physics, Lancaster University, LA1 4YB, United Kingdom
}%

\date{\today}

\begin{abstract}
Optical chips for quantum photonics are cutting-edge technology, merging photonics and quantum mechanics to manipulate light at the quantum level. These chips are crucial for advancing quantum computing, secure communication, and precision sensing by integrating photonic components like waveguides, beam splitters, and detectors to manipulate single photons, the fundamental carriers of quantum information.

Key advancements in optical chips include low-loss waveguides, efficient single-photon sources, and high-fidelity quantum gates, all essential for scalable quantum circuits. Integrating these circuits on a chip offers significant advantages in miniaturization, stability, and reproducibility over traditional bulk optics setups.

Recent breakthroughs in materials science and nanofabrication have propelled the field forward, enabling the production of chips with higher precision and lower defect rates. Silicon photonics, in particular, has become a prominent platform due to its compatibility with existing semiconductor manufacturing processes, facilitating the integration of quantum photonic circuits with classical electronic systems.

Here, we share our vision of the future of optical chips for quantum photonics, which hold promise for various applications. In quantum computing, they enable the development of compact and scalable quantum processors. In communication, they provide the foundation for ultra-secure quantum networks through quantum key distribution. In sensing, they allow for high-precision measurements that surpass classical limits. As research progresses, optical chips are expected to play a critical role in realizing the full potential of quantum technologies.
\end{abstract}

\maketitle

\section{\label{sec:level1}Introduction}

Optical chips for quantum photonics are emerging as a transformative technology at the intersection of photonics and quantum mechanics. These integrated platforms are designed to manipulate and control light at the quantum level, offering unprecedented capabilities for a wide range of applications, including quantum computing, secure communication, and precision sensing. By integrating various photonic components-such as waveguides, beam splitters, and detectors-onto a single chip, these devices enable the efficient manipulation of single photons, which are essential carriers of quantum information.

The development of optical chips for quantum photonics addresses several critical challenges in the field of quantum technologies. Key advancements include the fabrication of low-loss waveguides, the creation of efficient single-photon sources, and the realization of high-fidelity quantum gates. These components are crucial for constructing scalable quantum circuits, which are fundamental for practical and reliable quantum computing systems. Additionally, integrating quantum photonic circuits on a chip offers significant advantages in terms of miniaturization, stability, and reproducibility, surpassing the capabilities of traditional bulk optics setups.

Recent breakthroughs in materials science and nanofabrication techniques have significantly advanced the field, enabling the production of chips with higher precision and lower defect rates. Silicon photonics, in particular, has emerged as a prominent platform due to its compatibility with existing semiconductor manufacturing processes. This compatibility facilitates the integration of quantum photonic circuits with classical electronic systems. This compatibility paves the way for more complex and efficient hybrid systems essential for advancing quantum technologies.

Optical chips for quantum photonics hold promise for a multitude of applications. They offer a pathway to building compact and scalable quantum processors in quantum computing. They lay the foundation for ultra-secure quantum networks through quantum key distribution in communication. Moreover, in sensing, they enable high-precision measurements that surpass classical limits. As research and development in this field continue to progress, optical chips are expected to play a crucial role in realizing the full potential of quantum technologies, driving innovation and opening new frontiers in science and technology.

Quantum mechanics, derived from the Latin term 'quantum' meaning 'quantity', is a fundamental physics theory that describes particle behavior at atomic and subatomic scales. It has been pivotal in explaining phenomena in various fields, such as atomic structure and superconductivity. This perspective provides a forward-looking view of optical chips for quantum photonics, presenting unique insights, innovative ideas, and emerging trends within a particular field.

\section{Quantum mechanics history}

In the late 19th century, several unresolved questions emerged, challenging the foundations of classical physics. One major issue was the stability of atoms. According to Maxwell's theory and Hertz's experiments, accelerating charges emit radiation. Consequently, orbiting electrons should continuously lose energy and spiral into the nucleus, yet atoms were observed to be stable. Another significant problem was understanding the emission of specific spectral lines in gas spectroscopy. Although experiments confirmed these lines, their underlying mechanism remained unexplained. The photoelectric effect, discovered by Heinrich Hertz in 1887, further complicated the classical understanding. Hertz observed that ultraviolet (UV) light increased the conductivity of a material, but this effect was independent of the light's intensity below a specific frequency. This contradicted classical predictions that intensity should influence conductivity. Additionally, the black body radiation problem, addressed by Lord Rayleigh and James Jeans in 1900, revealed discrepancies between theoretical models and observed behaviors at shorter wavelengths, leading to what was known as the 'UV catastrophe.'

Addressing these problems through several key discoveries led to the formulation of the 'old quantum theory,' which paved the way for the development of modern quantum mechanics.

The first fundamental discovery was the quantization of energy and light. In 1900, Max Planck developed a theory of blackbody radiation that matched experimental results by proposing that the oscillations producing the waves are discrete and that the radiation is quantized into distinct units of energy \cite{planck1901law}. In 1905, Albert Einstein showed that the photo-electric effect, which was observed by Hertz, can be explained via the quantization hypothesis presented by Plank \cite{einstein1905erzeugung}. It is worth noting that Einstein was describing photon, the term 'photon' was only first introduced in 1916 by Leonard T. Troland as '\textit{A photon is that intensity of illumination upon the retina of the eye.}' \cite{troland1917measurement}. Later in 1926, Gilbert N. Lewis used the term photon as the unit of radiant energy - '\textit{I, therefore, take the liberty of proposing for this hypothetical new atom, which is not light but plays an essential part in every process of radiation, the name photon}' \cite{lewis1926conservation}. This photon definition differs from Einstein's, which described it as light quantum (quanta). Ernest Rutherford and Niels Bohr solved the big problem of the stability of atoms. In 1911, Ernest Rutherford experimentally showed by bombarding thin gold foil with alpha particles that atoms have a tiny nucleus with electrons cloud around it \cite{rutherford1911scattering}. In 1913 by adopting Rutherford's nuclear structure to Max Planck's quantum theory, Niels Bohr presented his model of the atom \cite{bohr1913constitution} according to which electrons are around the nucleus in a discrete set of orbits. The emission or the absorption of light occurs when an electron jumps between the orbits.

The work of Plank, Einstein and Bohr (known as the 'old quantum theory') was the key to developing the theory of quantum mechanics, known as the 'new quantum theory'. In 1924, Louis De Broglie extended the theory of wave-particle duality from light to matter \cite{de1923waves}, which became the base of wave mechanics. He proposed that each moving particle has an associated wave with a wavenumber proportional to the particle's momentum which he called a 'phase wave'. In 1927, his theory was confirmed by Davisson-Germer experiments, showing a diffraction pattern of reflected electrons by crystals \cite{davisson1927scattering,davisson1927diffraction,davisson1928reflection}. In 1925, a new quantum theory was developed/presented. In 1925, Schrodinger described a quantum system by wave mechanics when the state of the quantum system is quantified by the wave function \cite{schrodinger1926quantisierung}. The square of the wave function gives the probability density of finding a particle at a specific place at a specific time. In parallel to Schrodinger's theory, Werner Heisenberg, with Max Born and Pascual Jordan, presented matrix mechanics for describing quantum systems \cite{heisenberg1925quantum,born1925quantenmechanik,born1926quantenmechanik}. The first consistent quantum mechanics unifying both contradicting views was developed by Paul Dirac. In 1925 after seeing Heisenberg's paper, Dirac developed a theory for non-commuting observables described by linear operators on Hilbert spaces and the bra-ket notation, which is still used these days \cite{dirac1926theory}. In 1928, he presented the Dirac equation \cite{dirac1928quantum}, a relativistic generalization of the Schrodinger equation, adding the new variable spin, presented by Wolfgang Pauli in 1924. These discoveries were the foundations of the modern quantum physics.


\section{\label{sec:level2}From free-space to chip}

\begin{figure*}
    \includegraphics{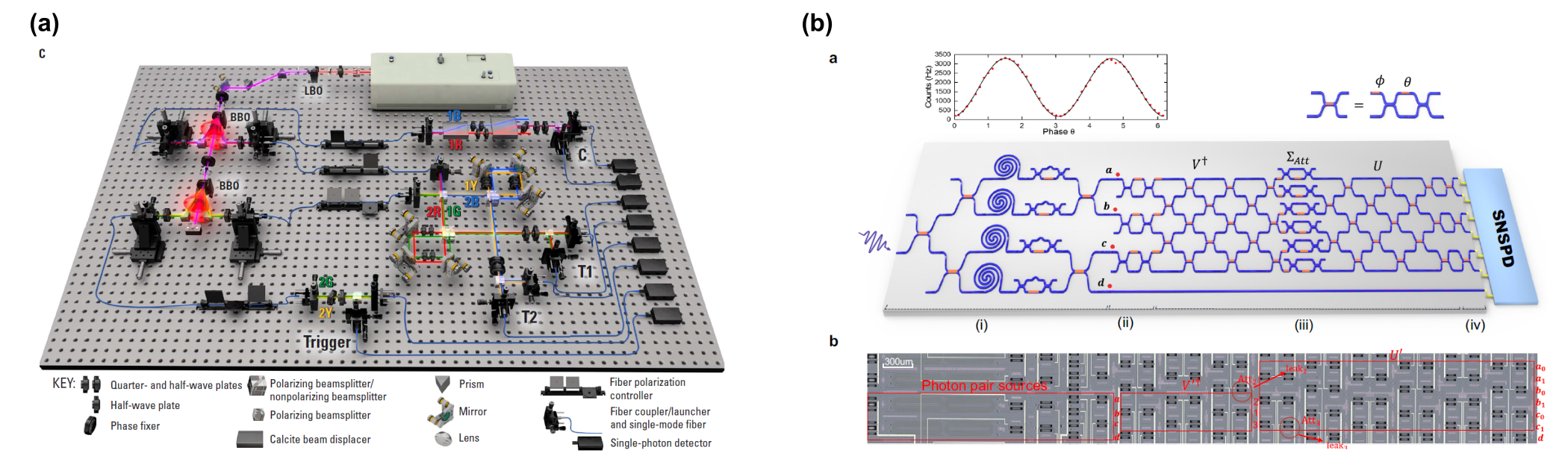}
    \caption{Optical logic gates circuits. (a) Optical Fredkin quantum gate (reproduced from \cite{patel2016quantum}). (b) Quantum logic gates on a chip (reproduced from \cite{li2022quantum}).}
    \label{fig:Figure1}
\end{figure*}

Quantum optics is a branch of quantum physics that studies light-matter interactions on the level of individual photons. Photons are excellent candidates for quantum information processing due to their high mobility and low interaction with the environment. Various applications are based on quantum effects. Conventional quantum optical systems are performed in free space on an optical setup with bulky and discrete optical components. Complex setups often require hundreds of components, necessitating a large optical table and challenging system deployment. Additionally, these setups require precise alignment and strict control over noise factors such as temperature and vibrations. Integrated photonics provides a solution to these problems for quantum technologies. Integrated quantum photonics uses photonic integrated circuits on a chip to control photonic quantum states for applications in quantum technologies. As such, integrated quantum photonics offers the only viable approach to the miniaturization and scaling up of optical quantum circuits. Figure~\ref{fig:Figure1} compares quantum circuits implemented in a free space setup and a chip. It shows that a conventional quantum circuit setup (Fig.~\ref{fig:Figure1}a) requires a bulky and robust optical table while a quantum circuit on a chip (Fig.~\ref{fig:Figure1}b) can have a length of a few millimeters.

Optical waveguides can control the light in microscale dimensions. The basic structure of an optical waveguide is a substrate and a guiding layer \cite{katiyi2017figure}. The light is guided by total internal reflection and propagating in the guiding medium. The first presentation of an optical guided wave structure was in 1966 when a cylindrical fiber was proposed for guiding light \cite{kao1966dielectric}. Since then, optical waveguides have been used for a variety of applications \cite{karabchevsky2020chip}, from passive components (such as directional couplers and sensors) to active components (such as light sources and modulators). Optical waveguides offer several advantages over bulk optics for quantum technology, including miniaturization, integrability, scalability, stability, and cost-effective production. Fabricating quantum devices on a chip makes them smaller and more stable. Optical waveguides can be made from various materials, both active and passive, to suit specific quantum components and applications. All necessary optical components can be integrated into a single chip, creating a robust circuit that reduces errors caused by vibrations or temperature fluctuations. Additionally, this approach lowers the cost of mass production.


\section{\label{sec:level3}Quantum technology devices}

\subsection{\label{sec:level4}Materials platforms for quantum technology}

Materials for on-chip nanophotonics are critical to the performance and functionality of photonic circuits \cite{katiyi2023passive}. Silicon has long been the foundation of on-chip photonics, thanks to its compatibility with semiconductor manufacturing \cite{orcutt2012open,kita2018high} and its ability to form high-quality, low-loss waveguides \cite{cardenas2009low}. However, its limitations, such as an indirect bandgap, have led to exploring alternative materials. Silicon nitride (Si$_3$N$_4$) is notable for its lower propagation losses and effectiveness in the visible and near-infrared spectra \cite{bose2024anneal,ji2023ultra,subramanian2013low,sacher2019visible}. III-V semiconductors, like gallium arsenide (GaAs) \cite{verrinder2021gallium,chang2018heterogeneously} and indium phosphide (InP) \cite{wang2024scaling,smit2019past}, are known for their direct bandgap properties, allowing efficient light emission and detection, making them ideal for active devices such as lasers and modulators. Emerging materials, including lithium niobate (LiNbO$_3$) \cite{qi2020integrated,yu2019coherent,cai2019acousto,wang2018integrated} and two-dimensional materials like graphene \cite{xia2009ultrafast,kovacevic2018ultra,ding2017efficient,guo2020high} and transition metal dichalcogenides (TMDs) \cite{liu2021mxenes,zhang2022mxenes,wang20232d}, bring unique electro-optic, nonlinear, and plasmonic properties, enhancing the capabilities of on-chip photonic systems. Research into hybrid platforms that integrate these diverse materials aims to combine their strengths, enabling more versatile and high-performance nanophotonic devices. Advances in materials science and nanofabrication are driving this field forward, opening up new possibilities for quantum computing, telecommunications, and other cutting-edge applications.

Quantum photonic circuits comprise several key building blocks, as previously mentioned. Quantum-integrated photonics combines different components, such as light sources and detectors, onto a single chip. Optical waveguides can be fabricated from monolithic or hybrid materials, making them ideal configurations for quantum technologies. Table~\ref{tab:table1} lists standard materials used in quantum devices, each with unique properties suited to various quantum applications. Different components may require different materials depending on their specific function. For instance, the silicon-on-insulator is CMOS-compatible and can be used for photon-pair generation via four-wave mixing \cite{matsuda2012monolithically,silverstone2014chip} but is most efficient for single-photon detection at 1550 nm. Monolithic circuits, made from a single material, such as III-V semiconductors like gallium arsenide (GaAs) and indium phosphide (InP), enable the integration of multiple components, including light sources and detectors, on one chip.

\begin{table*}
\caption{Common materials for quantum devices.}
\label{tab:table1}
\begin{ruledtabular}
\begin{tabular}{lllll}
    Material & Transparency & Non-linear index & Refractive index & Waveguide losses \\
    ~ & ~ & ~ & ($\lambda=1.55~{\mu}m$) & ~ \\
    \hline\hline
    Silica & VIS-IR & Weak $\chi^{(3)}$ & 1.44 & Ultralow  \\
    \hline
    Silicon on insulator & IR & Strong $\chi^{(3)}$ & 3.44 & Moderate linear loss, \\
    (SOI) & $>1,000$ nm & $\chi^{(3)}=6.5E-14$ & ~ & high two-photon absorption \\
    \hline
    Silicon nitride & VIS-IR & Strong $\chi^{(3)}$ & 1.99 & Low linear loss, \\
    (Si$_3$N$_4$) & 400-2400 nm & $\chi^{(3)}=2.5E-15$ & ~ & low two-photon absorption \\
    \hline
    Gallium arsenide & IR & Strong $\chi^{(2)}$ & 2.9-3.4 & Moderate \\
    (AlGaAs) & $>900$ nm & $d(1/2\cdot\chi^{(2)})=119$ & ~ & ~ \\
    \hline
    Lithium niobate & VIS-IR & Strong $\chi^{(2)}$ & 2.2 & Moderate  \\
    (LNOI) & 400-5000 nm & $d(1/2\cdot\chi^{(2)})=27$ & ~ & ~ \\
    \hline
    Indium phosphide & IR & Strong $\chi^{(2)}$ & 3.15 & Moderate \\
    (InP) & 900-2000 nm & ~ & ~ & ~ \\
    \hline
    Aluminium nitride & UV-IR & Moderate $\chi^{(2)}$ & 2.12 & Moderate \\
    (AlN) & 210-4000 nm & $\chi^{(2)}=4.7$ & ~ & ~ \\
\end{tabular}
\end{ruledtabular}
\end{table*}

\subsection{Single-photon sources}

\begin{figure*}
    \includegraphics{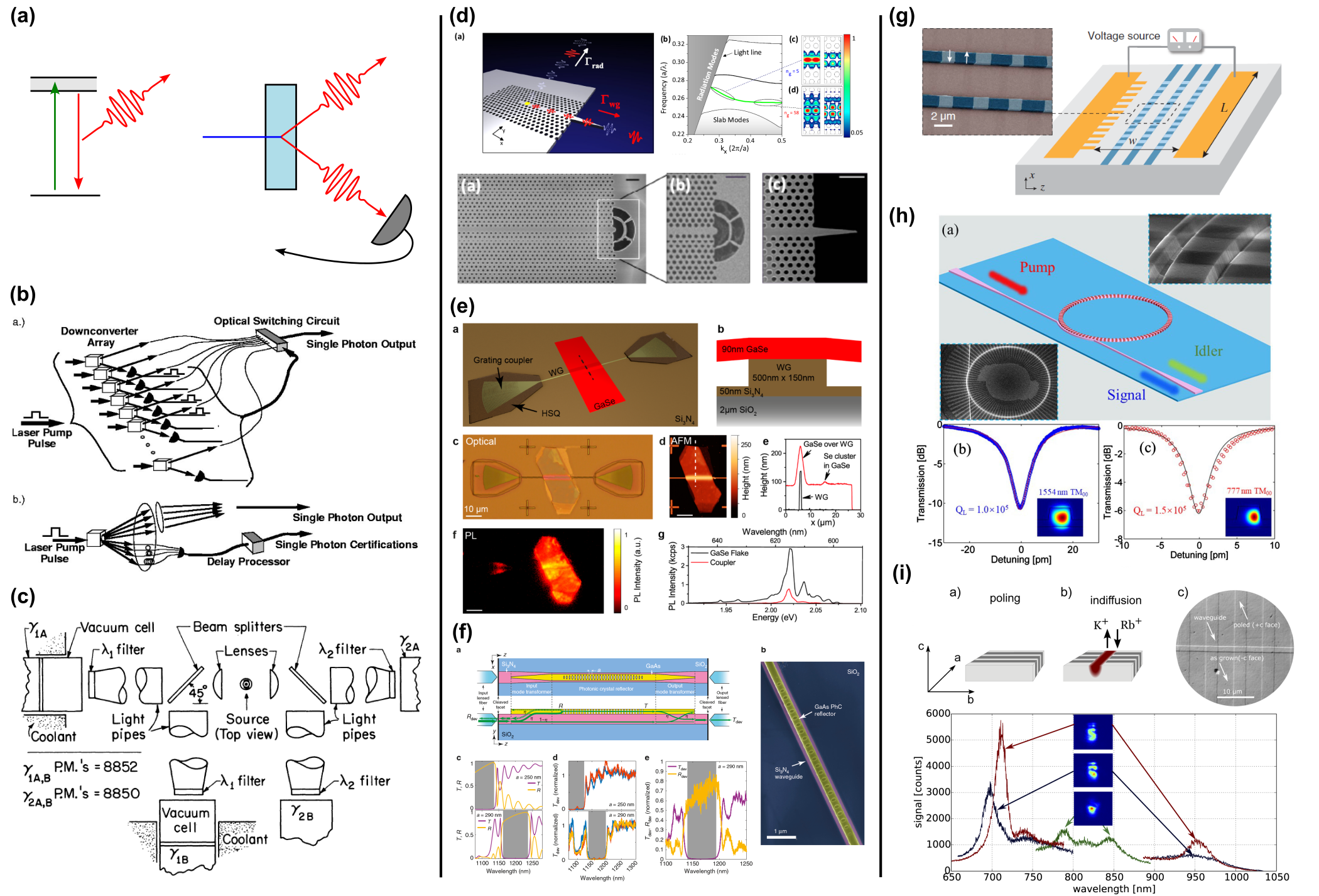}
    \caption{Single photon sources. (a) Illustration of the emission process of (left) single photon source and (right) photon pair source. (b) Illustration of probabilistic source multiplexing of many photon pair sources and multiplexing a single-photon source (reproduced from \cite{migdall2002tailoring}). (c) Schematic of the first free-space single photon source (reproduced from \cite{clauser1974experimental}). (d) A quantum dot single-photon source embedded in a photonic crystal waveguide  (reproduced from \cite{arcari2014near}). (e) Single-photon source of gallium selenide crystals on a silicon nitride rib waveguide (reproduced from \cite{tonndorf2017chip}). (f) Single-photon source of III-V semiconductor GaAs with Self-assembled InAs/GaAs QD heterogeneous fabricated on a silicon nitride waveguide (reproduced from \cite{davanco2017heterogeneous}). (g) Illustration of the periodically poling process of a nonlinear crystal waveguide (reproduced from \cite{wang2018ultrahigh}). (h) Periodically poled lithium niobate single photon source waveguide at a wavelength of 1550 nm (reproduced from \cite{ma2020ultrabright}). (i) Periodically poled Rb:KTP single photon source waveguide at a wavelength of 800 nm (reproduced from \cite{eigner2020spatially}).}
    \label{fig:SPS}
\end{figure*}

The concept of a single-photon source was first introduced by Planck in 1900. A single-photon source emits one photon at a time with a known and identical mode, such as frequency or temporal mode, which is crucial for quantum technologies. However, an ideal single-photon source does not yet exist, and ongoing efforts aim to create sources that closely approximate this ideal. Two primary methods for generating single-photon sources are isolated quantum systems (also known as single-emitter systems) and photon pair sources, as shown in Fig.~\ref{fig:SPS}a. Another method for creating single-photon sources is by multiplexing one or more photon pair sources, as illustrated in Fig.~\ref{fig:SPS}b.

Isolated quantum systems are deterministic sources that emit one photon at a fixed time but with an efficiency of less than 100\%. For example, quantum dots \cite{flissikowski2001photon,buckley2012engineered,he2013demand}, single ions \cite{keller2004calcium,barros2009deterministic,muller2015single} and single atoms. Notably, quantum dots can be used in both isolated quantum systems and photon pair sources. The second method, photon pair sources, generates pairs of correlated photons, where the detection of one photon ('twin') indicates the presence of the other. Unlike isolated quantum systems, photon pair sources guarantee photon emission. However, since they rely on nonlinear processes, they are probabilistic in nature. Photon pair sources are typically based on two nonlinear effects: spontaneous parametric down-conversion (SPDC) and four-wave mixing (FWM). Parametric down-conversion is related to the second order susceptibility $\chi^{(2)}$ and can occur in crystals such as lithium niobate (LiNbO$_3$) and gallium arsenide (GaAs). In the SPDC process, one pump photon is split into signal and idler photons. Four-wave mixing is related to the third order susceptibility $\chi^{(3)}$ and is based on the atomic structure of centrosymmetric materials such as silicon oxide (SiO$_2$) and silicon nitride (Si$_3$N$_4$). In the FWM process, two pump photons are converted into a pair of signal and idler photons.

In 1974, the first source of entangled photons was created \cite{clauser1974experimental}. By splitting the first and second atoms of an atomic cascade in mercury atoms, as shown in Fig.~\ref{fig:SPS}c, they observed one photon per pulse. One limitation of free-space single-photon sources is their bulkiness. Integrated photonics allows the fabrication of small single-photon sources with chip-scale dimensions. Integrated quantum photonics can create high-brightness sources because of the high confinement in optical waveguides. Due to the high power confinement in optical waveguides, the non-linear effects, such as spontaneous parametric down-conversion (SPDC) and four-wave mixing (FWM), are enhanced. In addition, it decreases insertion losses that occur in the fiber/chip interface. A single photon source can be created using quantum dots and periodically poled nonlinear crystals (quasi-phase matching). Quantum dots are a good candidate for photon sources due to the near-ideal single-photon emission and entangled photon-pair generation. Quantum dot can be embedded in a photonic crystal waveguide \cite{laucht2012waveguide,arcari2014near} as shown in Fig.~\ref{fig:SPS}d. The photonic crystal waveguide can achieve a near-unity probability for a single exciton ($\beta={\sim}98\%$) that can be utilized for highly efficient single-photon source \cite{arcari2014near}. Quantum dot can also be integrated on a waveguide by layered semiconductor \cite{tonndorf2017chip} as shown in Fig.~\ref{fig:SPS}e. A layer of gallium selenide (GaSe) single crystal with a thickness of 90 nm was placed on Si$_3$N$_4$ rib waveguide for coupling a single-photon source to a waveguide. The integration can be improved by III-V semiconductor GaAs with Self-assembled InAs/GaAs QD heterogeneous fabricated on a silicon nitride \cite{davanco2017heterogeneous} as shown in Fig.~\ref{fig:SPS}f. The advantage of using GaAs is its well-established fabrication processes, which facilitate the fabrication of desired applications. For example, in the integrated silicon nitride waveguide, the GaAs layer can be designed and fabricated as a tapered waveguide for better mode conversion and as a photonic crystal reflector for high efficiency.

Another method for generating single-photon sources on a waveguide is quasi-phase-matching (QPM) nonlinear frequency conversion, achieved by periodically poling a nonlinear crystal \cite{hum2007quasi}. By applying an electric field on a material with ferroelectric properties, as illustrated in Fig.~\ref{fig:SPS}g, the signs of the elements in $\chi^{(2)}$ tensor are switched, which can be manipulated in the required configuration for QPM. One of the materials that can be periodically poled is lithium niobate (LN - LiNbO$_3$) \cite{shur2015micro}. Lithium niobate is an artificial ferroelectric crystal with strong $\chi^{(2)}$ nonlinearity. A quasi-phase matching is obtained by periodic poling along the propagation direction of the LN waveguide, which offsets the mismatch between the interacting waves as shown in Fig.~\ref{fig:SPS}g. The phase matching can be used for efficient nonlinear effects on a waveguide, such as second-harmonic generation \cite{wang2018ultrahigh}. This method was first introduced in 1993 for developing a blue laser using second-harmonic generation \cite{yamada1993first}. Figure \ref{fig:SPS}h shows a periodically poled lithium niobate microring waveguide for a photon pair source at a wavelength of 1.55 $\mu$m \cite{ma2020ultrabright}. Other materials for periodically poled waveguides are Potassium titanyl phosphate (KTiOPO$_4$, KTP) and its isomorphs (KTP family). Figure \ref{fig:SPS}i shows a photon pair source in periodically poled rubidium (Rb) exchanged Potassium Titanyl Phosphate (Rb:KTP) waveguide around 800 nm \cite{eigner2020spatially}. It achieved a high nonlinear conversion efficiency ($2\%/\text{W}\cdot\text{cm}^2$) and low-temperature sensitivity (39~\text{pm}/\text{K}), demonstrating excellent potential for sources in the visible spectrum. KTP presents unique advantages that could surpass LN in various high-power and visible-light applications. Notably, KTP’s lower photorefractive damage threshold and reduced coercivity fields make it a promising candidate for periodically poled waveguides. These properties open avenues for KTP's use in high-power environments and support the fabrication of dense quasi-phase-matching (QPM) gratings suitable for visible wavelengths—applications where lithium niobate (LN) faces limitations. However, challenges remain in optimizing KTP's crystal growth, which currently requires more costly techniques and yields smaller crystal sizes. Overcoming these hurdles could significantly expand KTP’s role in photonic device innovation.

As mentioned, a significant drawback of photon pair sources is that their generation process is probabilistic, which poses a problem for quantum applications. A way to overcome this drawback is to convert a probabilistic source into a deterministic source by multiplexing, as illustrated in Fig.~\ref{fig:SPS}b. In 2001, Clausen proposed the concept of multiplexing using an optical feedback loop \cite{clausen2001conditional}. In 2002, Pittman showed the first experimental demonstration of multiplexing with a periodic time interval using a storage loop and an optical switch \cite{pittman2002single}. One method for multiplexing is multiplexing many photon pair sources to create a deterministic single-photon source \cite{christ2012limits,tiedau2019scalability}. The important component of sources multiplexing is the optical switch. The optical switch, such as opto-ceramic switches \cite{xiong2013bidirectional} and electro-optic switches \cite{meany2014hybrid}, gets feedback from the heralding detector and switches the herald photon to the output. Another method is multiplexing a single-photon source in time or frequency \cite{migdall2002tailoring,joshi2018frequency}. A probabilistic source generates photons at random times, and the heralding detector collects the timing. The herald photons are introduced into a delay system, while the switch adjusts the repetition rate according to the recorded timing.

While single-photon sources can be integrated on a chip, efficient coupling of the high-power pump source to the waveguide remains essential to ensure sufficient excitation. The common method for enhancing the coupling efficiency is butt-coupling using a lensed fiber. Lensed fibers can focus the light to a spot size of 1 $\mu$m, which reduces the large mismatch between the waveguide dimensions and light spot size, increasing the coupling efficiency. Butt-coupling using a lensed fiber offers a straightforward and broadband-compatible approach, providing higher coupling efficiency than standard single-mode fibers. This method has been widely employed for interfacing various waveguide-integrated single-photon sources. \cite{collins2013integrated,ma2020ultrabright,davanco2017heterogeneous}. For achieving a higher coupling efficiency, a grating coupler can be fabricated on the waveguide, enhancing the coupling efficiency of the needed wavelength. Grating couplers operate based on the Bragg condition, making them inherently wavelength-selective and suitable only for a specific, pre-designed wavelength. It can achieve a coupling efficiency of $\sim$80\% \cite{sacher2014wide,marchetti2017high}. The grating can be in a tapered shape, focusing the light into the waveguide \cite{wang2014focusing,tonndorf2017chip}. Implementing a tapered grating coupler in a single-photon source waveguide can be seen in Fig. \ref{fig:SPS}e. It is worth noting that lensed fiber can be used for focusing the incident light into the grating in a small spot size, making the grating more compact. The coupler can be designed as a gradient grating or a metasurface, introducing distinctive propagation characteristics within the waveguide \cite{sapra2019inverse,gu2021optical,chen2024freeform,gong2017hybrid}. Another method that can be used is an edge coupler. It can achieve 90\% coupling efficiency and can support broadband operation \cite{bakir2010low}.

\subsection{Single photon detectors}

Detecting a single photon is a complex yet crucial task for advancing quantum technology. Integrating single-photon detectors onto waveguides enables miniaturization and reduces optical coupling losses when interfacing with fibers, paving the way for more efficient and compact quantum devices. Several requirements must be met to achieve an ideal single-photon detector. First, it must be susceptible and capable of detecting a single photon at 100\% efficiency. Second, a few crucial parameters of the detectors must equal zero: the dark count rate, the dead time (or recovery time), and the timing jitter. Another essential attribute of a single-photon detector is its ability to resolve the number of photons in a pulse. Single-photon detectors are divided into non-photon-number-resolving detectors and photon-number-resolving detectors.

\begin{figure*}
    \includegraphics{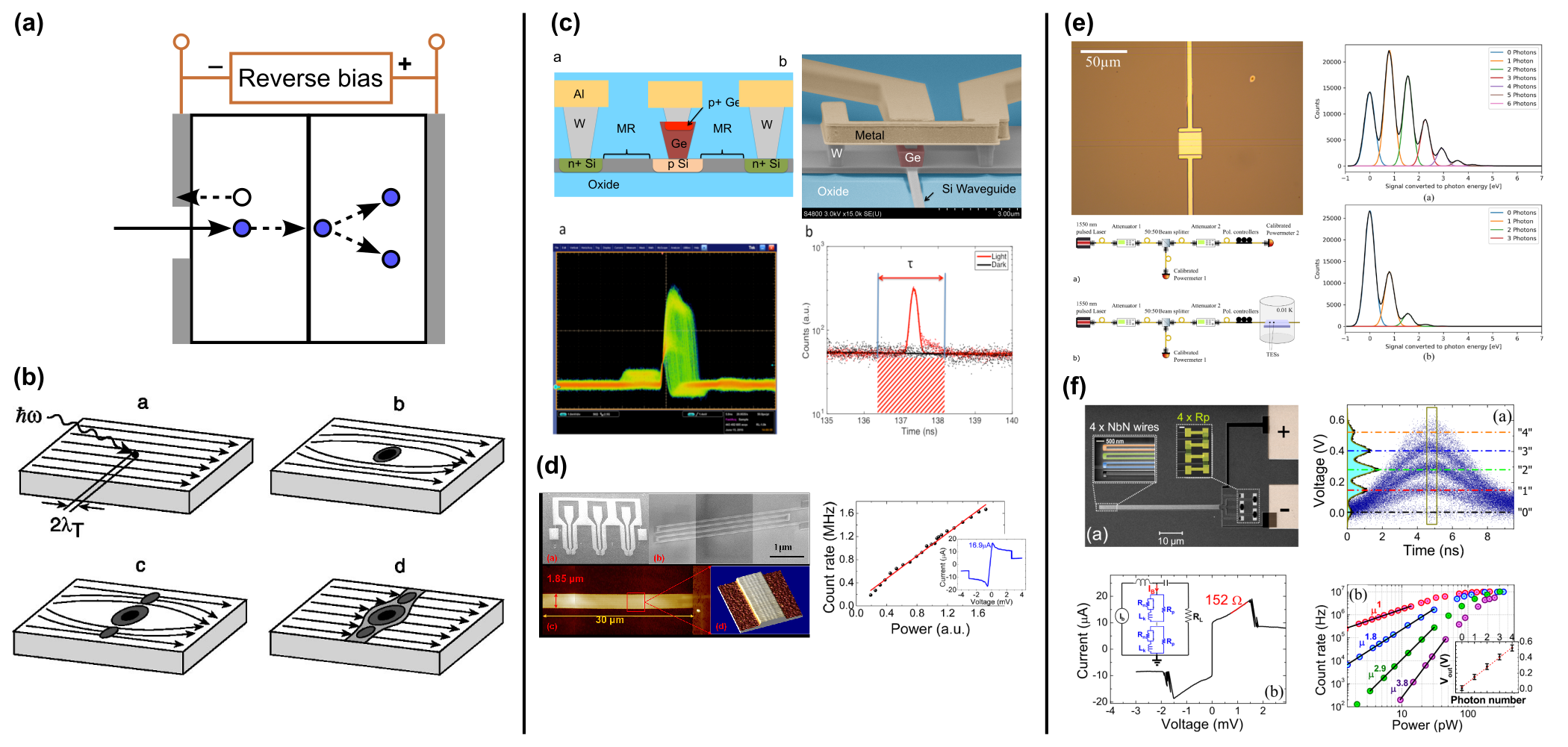}
    \caption{Single photon detectors. (a) Illustration of the concept of a single-photon avalanche photodiode (SPAD). (b) Illustration of the concept of superconducting nanowire single-photon detector (SNSPD) (reproduced from \cite{gol2001picosecond}). (c) Ge-on-Si lateral avalanche photodiode waveguide (reproduced from \cite{martinez2017single}). (d) Superconducting nanowire single-photon detector waveguide (reproduced from \cite{sprengers2011waveguide}). (e) Photon-number-resolving transition edge sensor integrated on a lithium niobate waveguide (reproduced from \cite{hopker2019integrated}). (f) Photon-number-resolving four niobium nitride (NbN) superconducting nanowires integrated on a GaAs waveguide  (reproduced from \cite{sahin2013waveguide}).}
    \label{fig:SPD}
\end{figure*}

The vast majority of single-photon detectors today are non-photon-number-resolving, distinguishing only between the presence and absence of photons rather than counting individual photons. Since the first demonstration of single-photon detection using the photomultiplier tube (PMT) in 1949, the quest for effective photon detection has continued to evolve. The PMT, though able to amplify the signal up to 10$^6$ electrons, operates with limited efficiency (10–14\%) and requires a vacuum tube, impacting its reliability, longevity, and size. Today, the single-photon avalanche diode (SPAD) or Geiger-mode avalanche photodiode (G-APD), conceived by Jun-ichi Nishizawa in 1952, has emerged as a highly practical alternative. By applying a high reverse bias just above the breakdown voltage in the p-n junction, SPADs utilize impact ionization, also known as the avalanche effect, to generate a self-sustaining current upon photon absorption. This Geiger-mode operation, introduced in the 1960s, presents a streamlined pathway for single-photon detection in compact, scalable designs, positioning SPADs as a cornerstone for advancing quantum technology and integrated photonics \cite{haitz1964model,haitz1965mechanisms}. SPADs are fast and have low noise, but have low quantum efficiency ($n_d \approx 65\%$). The advantage of SPAD is that it can operate at room temperature or slightly below it. In 2017, a lateral germanium (Ge) avalanche photodiode operating in Geiger mode was integrated on a silicon waveguide as a single-photon detector \cite{martinez2017single} as shown in Fig.~\ref{fig:SPD}c. It achieved single-photon detection with an efficiency of 5.27\% at an operation temperature of 80K.

The emergence of superconducting nanowire single-photon detectors (SNSPDs) marks a significant advancement in photon detection technology. These innovative devices have the potential to transform the landscape of quantum optics and photonics, offering enhanced sensitivity and efficiency that could pave the way for new applications and breakthroughs in quantum information science \cite{gol2001picosecond,rosfjord2006nanowire}. SNPADs can achieve higher efficiency ($>90\%$), low dark count rate ($<0.01~\text{cps}$) and high count rate from visible to near-infrared. Currently, a superconducting nanowire is applied close to the critical current. When even a single photon is absorbed, it contributes to the average resistivity. It greatly enhances the detection efficiency but requires a cryogenic temperature to operate. It was first presented in 2001 by a microbridge from niobium nitride films on sapphire substrate \cite{gol2001picosecond}. At a wavelength of 0.81 $\mu$m, they achieved quantum efficiency of 20\% for 0.81 $\mu$m photons, negligible dark counts and response time of $\sim$100 ps. A few years later, the efficiency increased to 67\% \cite{rosfjord2006nanowire} while reaching above 90\% efficiency \cite{marsili2013detecting}. In 2011, the first superconducting nanowire single photon detector on a waveguide was presented \cite{sprengers2011waveguide} as shown in Fig.~\ref{fig:SPD}d. Niobium nitride nanowires were fabricated on a GaAs ridge waveguide for sensing the evanescent field, achieving $\sim$20\% efficiency at the telecommunication range.

A key property of single-photon detectors is their photon-number-resolving (PNR) capability. It is essential when photon number counting is important in quantum protocols such as quantum computing and quantum communication. A transition-edge sensor (TES) can achieve PNR and be utilized for quantum applications \cite{cabrera1998detection}. It acts as a bolometer for single-photon detection. The temperature of a metallic detecting area is maintained at a superconducting state close to the ohmic state. The absorption of a photon causes a small rise in the temperature, which increases the resistance at a linear response, assuming that not too many photons are absorbed. The resistance is measured by an integrating circuit and the number of photons is calculated. TES was integrated on a lithium niobate waveguide, as shown in Fig.~\ref{fig:SPD}e, achieving a resolution of up to 6 photons at 1550 nm \cite{hopker2019integrated}. The main drawback of TES is a slow operation, two orders of magnitude higher than the needed speed for quantum computing \cite{calkins2011faster,lamas2013nanosecond}. Another solution is modifying an SNSPD for PNR quantum applications. By fabricating four niobium nitride (NbN) superconducting nanowires on a GaAs waveguide, as shown in Fig.~\ref{fig:SPD}f, a four-photon resolution was achieved \cite{sahin2013waveguide}. The main drawback is that both TES and SNSPD require cryogenic temperatures to operate, limiting their applications.

\subsection{Photon manipulation components}

At the heart of quantum technology lies the manipulation of photons, which possess various degrees of freedom -polarization, phase, spatial, spectral, and temporal- that can be controlled. Each dimension opens up unique avenues for encoding and processing quantum information. For instance, the polarization states of photons in a waveguide can be harnessed for polarization-encoded qubits, providing a robust framework for quantum computation and communication. To achieve this level of control, a diverse array of structures can be employed to manipulate these degrees of freedom. The efficacy of such manipulations is significantly enhanced by optical elements, such as polarization splitter-rotators\cite{ma2015symmetrical,dai2016realization}, directional couplers \cite{xu2019low}, phase shifters \cite{ikeda2010phase,harris2014efficient} and Mach-Zehnder interferometers, which can be integrated into optical waveguides, facilitating intricate photon management and fostering advancements in quantum technologies. The Pockels electro-optic effect or thermo-optic effect can be used to perform phase modulation. The Pockels effect, on the one hand, can be utilized for high-speed applications \cite{weigel2018bonded}, but is limited to non-centrosymmetric materials, such as lithium niobate (LiNbO$_3$) and gallium arsenide (GaAs). The thermo-optic effect, on the other hand, can be utilized for slow-speed applications \cite{harris2014efficient} and is readily achievable in silicon. Integrated photonics can achieve polarization modulation more effectively than free space. While free space requires bulky optical components, an optical waveguide can achieve polarization manipulation by a laser-written twisted waveguide \cite{morozko2023chip,sun2022chip}.


\section{\label{sec:level5}Applications and the Future directions}

Optical chips are revolutionizing quantum photonics by providing compact, scalable, and highly efficient platforms for manipulating the quantum states of light. Unlike conventional bulk optical systems, these chips integrate intricate optical circuits onto a single chip, allowing precise photon control while significantly reducing noise and environmental interferences. This breakthrough has catalyzed a wide array of applications, including quantum communication, cryptography, ultra-sensitive sensing, and quantum computing.

As the technology evolves, optical chips are emerging as pivotal components for next-generation quantum technologies. This chapter explores the transformative applications of optical chips in quantum photonics, highlighting innovations in integrated photonic circuits, single-photon sources, and quantum gate operations. By enabling scalable, reliable, and commercially viable solutions, optical chips are unlocking new possibilities for advancing quantum systems and reshaping the technological landscape.\\

\begin{figure*}
    \includegraphics{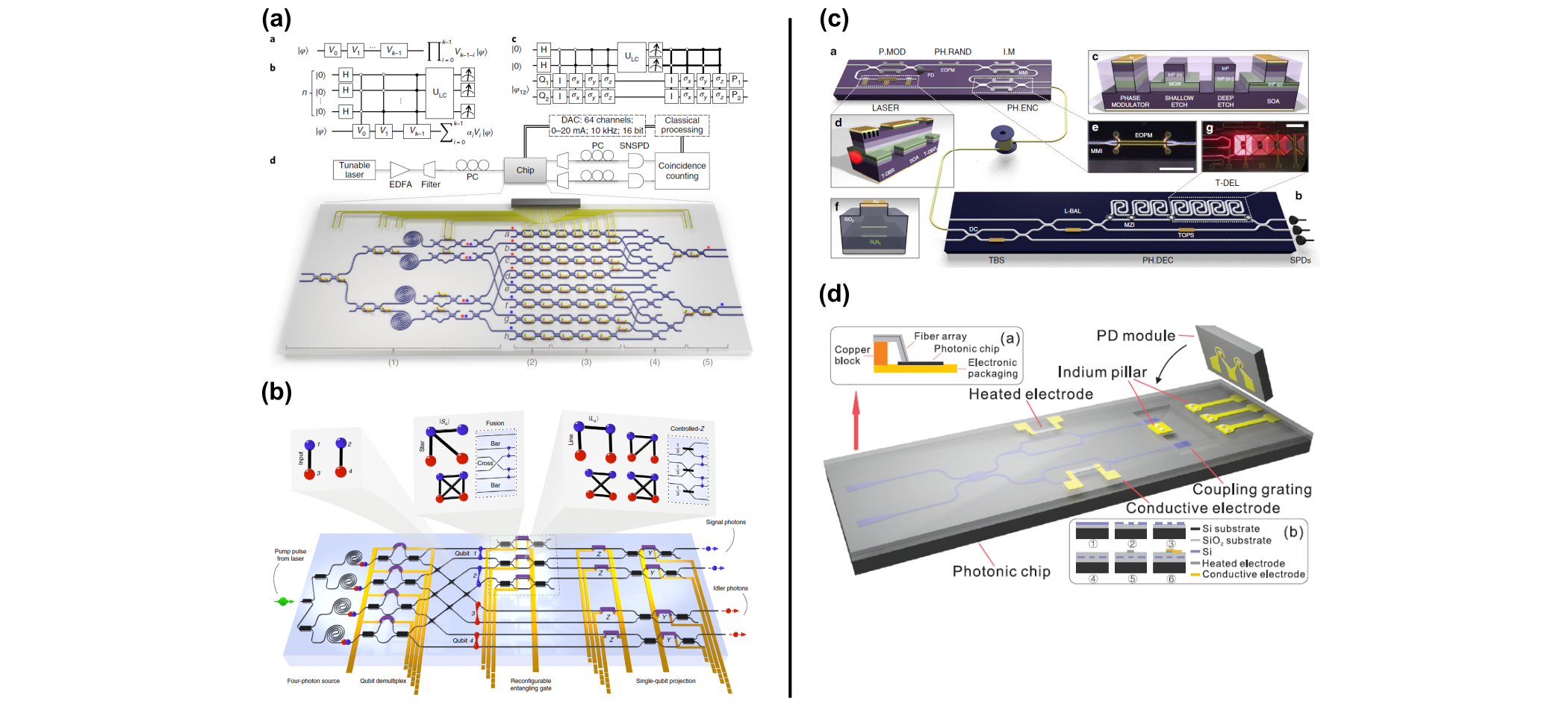}
    \caption{Quantum technology on a chip. (a) Programmable two-qubit quantum processor for quantum information processing on a chip (reproduced from \cite{qiang2018large}). (b) Programmable four-photon graph states on a silicon-on-insulator chip (reproduced from \cite{adcock2019programmable}). (c) Quantum key distribution (QKD) with low error rate on integrated indium phosphide chip (reproduced from \cite{sibson2017chip}). (d) Integrated hybrid quantum random number generator (QRNG) with a high generation rate on a silicon-on-insulator chip (reproduced from \cite{bai202118}).}
    \label{fig:Apps}
\end{figure*}

Figure~\ref{fig:Apps} shows a few quantum technologies implemented on a chip.

\subsection{Quantum computing and simulators}

Quantum computing holds great potential to enhance classical computing and can be leveraged to solve complex problems, including those in data science and quantum chemistry. It exploits quantum phenomena such as entanglement and superposition to achieve higher computational capabilities. Mathematician Yuri Manin first proposed quantum computing in 1980 \cite{manin1980computable}, and a year later, Richard Feynman proposed it \cite{feynman2018simulating}. Feynman declared that quantum system couldn't be simulated with the classical system; '\textit{Nature isn't classical, dammit, and if you want to make a simulation of nature, you'd better make it quantum mechanical, and by golly, it's a wonderful problem because it doesn't look so easy}'. 

Photons are light particles that can be manipulated and controlled to carry and process quantum information. In photonics-based quantum computing, qubits are encoded in the quantum states of photons and serve as the basic components of quantum computing. As opposed to binary code-based machines where a bit can only have values of 0 and 1, a quantum bit (or qubit) can store information in the superposition of states; $|\psi\rangle=\alpha|0\rangle+\beta|1\rangle$ where $\alpha$ and $\beta$ are complex numbers ($|\alpha|^2+|\beta|^2$=1). Superposition enables quantum computing to tackle problems with high computational complexity. Each additional qubit doubles the computational space, resulting in an exponential increase that reaches $2^n$ states. In classical computing, achieving the same computational space requires doubling the number of bits, resulting in significantly slower growth in capacity. The two most widely used methods for encoding qubits in photons are polarization encoding and path encoding. Quantum logic gates, such as the Controlled-NOT (CNOT), Hadamard (H), and Toffoli gates, are crucial for manipulating quantum information. The first quantum logic gate integrated on a chip was introduced in 2008, marking a key milestone in the development of quantum photonics on integrated platforms \cite{politi2008silica}.

Quantum simulators are an important application of quantum computers. Quantum simulators simulate and solve complex computational problems by using entanglement and many-particle quantum phenomena. The first quantum simulator was in 2002. A Rb Bose-Einstein condensate was used to observe a quantum phase transition \cite{greiner2002quantum}.

\subsection{Quantum sensing and metrology}

Quantum sensing and metrology are among the most promising fields within quantum information science. It is the measurement of physical quantities using quantum mechanical laws for sensitivity beyond the classical limit, such as atomic clocks \cite{ludlow2015optical}, nuclear magnetic and paramagnetic resonances \cite{boss2016one,simpson2017electron}, and electron microscopes. For example, classical measurements with a clock can achieve time resolution of femtoseconds (10$^{-15}$ seconds) to attoseconds (10$^{-18}$ seconds). Modern atomic clocks, particularly optical lattice clocks and ion-based atomic clocks, achieve resolutions far beyond those of classical clocks. State-of-the-art atomic clocks, such as those based on strontium or ytterbium in optical lattice setups, can achieve timing resolution on the order of attoseconds. This is achieved by exploiting atomic transitions in the optical range, where higher frequencies allow for finer time division. Quantum sensing can probe temperature, electric and magnetic
fields, forces, etc. Various research areas can benefit from the enhanced sensitivity of quantum sensing, including biological systems, spectroscopy and frequency measurements, gravitational wave detection, and magnetometry. 

Phase estimation allows quantum systems to achieve measurement precisions that surpass the standard quantum limit (SQL), which bounds classical measurement accuracy due to quantum noise. Therefore, phase estimation is highly preferable and widely used in quantum metrology. This technique is foundational for achieving high precision in measurements at the quantum level, as it enables the detection of extremely small changes in physical quantities. The super-resolution for phase estimation is more accessible and less demanding. The most common method to estimate the phase shift between two optical paths is the Mach-Zehnder interferometer (MZI)\cite{pezze2006phase,pezze2008mach}.

Integrated circuits enable scalable experimental platforms, providing control over noise effects such as temperature and vibrations, which facilitates precise phase control that is challenging to achieve in bulk optical setups. A promising platform for quantum sensing is the photonic crystal, which can be engineered with a photonic bandgap \cite{oudich2023tailoring,wu2022photonic}. Solid dielectric high-Q cavities, such as disk resonators\cite{mirzapourbeinekalaye2022free}, ring resonators\cite{belsley2022advantage}, and Mie nanoparticle resonators\cite{ivriq2025enhancing,kuznetsov2022special}, offer another viable platform. These structures can be used to implement single-photon-controlled gates.

Quantum atomic clocks have enabled timekeeping to reach the attosecond range, achieving resolutions and stabilities far beyond classical limitations. Their accuracy, measured in parts per quintillion, is foundational to advancing fields like navigation, telecommunications, and fundamental physics.

\subsection{Quantum cryptography}

Quantum cryptography utilizes quantum mechanics for cryptographic tasks such as quantum key distribution (QKD), quantum money and random number generator (RNG) to name a few. Photons have less interaction with the environment, saving the coded information. The fundamental concept behind Quantum cryptography is the no-cloning theorem. Quantum cryptography utilizes photons in quantum entanglement, where two photons are linked to each other. In 1982, Wootters and Zurek demonstrated that it is impossible to clone a quantum system or state \cite{wootters1982single}. When one attempts to measure a single photon, a change occurs in the other photon, preventing the information from being cloned or read.

\subsubsection{Quantum key distribution - private key and public key}

In the 1960s, the idea of Quantum cryptography (private key) was first proposed by Stephen Wiesner. He wrote a paper in 1968 but didn't publish it till 1982 \cite{wiesner1983conjugate}. Later in 1984, building upon Stephen Wiesner’s pioneering concept of quantum money and conjugate coding, Charles Bennett and Gilles Brassard introduced a groundbreaking protocol for quantum key distribution (QKD), known as BB84. This protocol leveraged the fundamental principles of quantum mechanics—such as the no-cloning theorem and the Heisenberg uncertainty principle—to enable two parties to securely generate a shared cryptographic key while ensuring that any eavesdropping attempt would be detectable. By encoding information in the polarization states of single photons, BB84 established the foundation for modern quantum cryptography, laying the groundwork for secure communication resistant to computational attacks.

Security relies on the use of an unpredictable key, which is generated by quantum random number generators (QRNGs). Unlike pseudo-random number generators, which follow deterministic patterns, quantum random number generators (QRNGs) produce genuinely random numbers. The integration of quantum photonics technology significantly reduces device size, making QRNGs more compact and practical for secure applications.

\textbf{In summary}, optical chips for quantum photonics represent a transformative technology that integrates photonics and quantum mechanics to manipulate light at the quantum level. These chips are essential for advancing quantum computing, secure communication, and precision sensing by enabling the controlled manipulation of single photons. Key developments, including low-loss waveguides, efficient single-photon sources, and high-fidelity quantum gates, have significantly enhanced scalability and performance. Recent advances in materials science and nanofabrication, particularly in silicon photonics, have further enhanced chip precision and integration with classical electronics. As research progresses, optical chips are poised to revolutionize quantum technologies, enabling compact quantum processors, ultra-secure communication networks, and highly precise measurement systems.




\section*{Data Availability Statement}

Data sharing is not applicable to this article as no new data were created or analyzed in this study.

\section*{Author declarations}

\subsection*{Conflict of Interest}
The authors have no conflicts to disclose.

\nocite{*}
\bibliography{aipsamp}

\end{document}